\documentclass[twocolumn,showpacs,prb,amsmath,amssymb]{revtex4}
\usepackage{graphicx}
\usepackage{dcolumn}
\usepackage{bm}

\begin{document}

\preprint{}

\title{Anharmonic effects in MgB$_2$? \\
A comparative inelastic X-ray scattering and Raman study}

\author{Matteo d'Astuto$^1$, Matteo Calandra$^1$, Stephanie Reich$^2$,
Abhay Shukla$^1$,
Michele Lazzeri$^1$,
Francesco Mauri$^1$,
Janusz Karpinski$^3$,
N.D. Zhigadlo$^3$,
Alexei Bossak$^4$, and
Michael Krisch$^4$.}
\affiliation{
$^1$ IMPMC, Universit\'es Paris 6 et 7, CNRS, IPGP, 140 rue de Lourmel, 75015 Paris, France. \\
$^2$ Department of Materials Science and Engineering,
Massachusetts Institute of Technology , 77 Massachusetts Avenue,
Cambridge, Massachusetts 02139-4307, USA\\
$^3$ Laboratory for Solid State Physics, ETH Zürich, 8093-Zürich, Switzerland\\
$^4$ European Synchrotron Radiation Facility, BP 220, F-38043 Grenoble cedex, France}

%
%
%
%

\date{\today}

\begin{abstract}
We study anharmonic effects in MgB$_2$ by comparing
Inelastic X-ray and Raman scattering together with {\it ab-initio }
calculations. Using
high statistics and high q resolution measurements 
we show that the E$_{2g}$
mode linewidth is independent of temperature along $\Gamma$A.
We show, contrary to previous claims, that 
the Raman-peak energy decreases as a function of 
increasing temperature, a behaviour inconsistent with all the 
anharmonic ab-initio calculations
of the E$_{2g}$ mode at $\Gamma$ available in literature. These
findings and the excellent agreement between the X-ray measured
and {\it ab initio} calculated phonon spectra
suggest that anharmonicity is not the main mechanism
determining the temperature behaviour of the Raman-peak energy.
The Raman E$_{2g}$  peak position
and linewidth can be explained by large dynamical effects in the 
phonon self-energy.
In light of the present findings, the commonly accepted explanation of
the reduced isotope effect in terms of anharmonic effects needs to be
reconsidered. 

\end{abstract}


\maketitle

\section{\label{intro}Introduction}

The discovery of MgB$_2$\cite{nagamatsu} demonstrated that
electron-phonon mediated superconductivity can generate 
critical-temperatures (T$_c=39$K) similar to those of perovskite oxides
superconducting systems such as cuprates \cite{bm} and doped {$\mathrm{BaBiO_{3}}$}
\cite{babio}.  Almost as surprising as the 39 K critical temperature
is the reduced isotope effect \cite{HinksNature,Budko}. The
partial isotope coefficients $\alpha(X)$, defined as
$\alpha(X)=-\frac{d\ln T_c}{d\ln M_X}$ where $M_X$ is the atomic mass
of ion $X$, are $\alpha(B)=0.30(1)$ and $\alpha({\rm Mg})=0.02(1)$.
Thus, the total isotope coefficient of $\alpha\approx 0.32$ is
substantially reduced from the canonical BCS value of $\alpha_{BCS}= 0.5$.

Deviations from $\alpha_{BCS}$ value are usually attributed to strong
electron-phonon coupling and to Coulomb repulsion.  However,
theoretical investigations of the isotope coefficient using single- or
double-gap Migdal-Eliashberg equations \cite{choi1,choi2} lead to
unsatisfactory results since the isotope effect reduction is
substantially underestimated.  For this reason, it was
proposed\cite{liu,kortus,choi1,choi2} that the T$_c$ in MgB$_2$ is
determined by the interplay of electron-phonon coupling and anharmonic
effects.  The latter can reduce T$_c$ through a hardening of
the phonon frequencies.  Refs.\cite{choi1,choi2} showed
than a hardening of $\approx$ 25\% of the E$_{2g}$ phonon-frequency,
along the high symmetry direction $\Gamma$A,
would be necessary to explain the reduced isotope-coefficient.

Anharmonic effects contribute to the broadening of a phonon and determine
energy shift from its harmonic value.
The anharmonic shift can be
determined from the actual experimental phonon energy and from the
harmonic energy, obtained from {\it e.g.} calculations.  The intrinsic
linewidth of a phonon is, in general, 
given by two contributions due to anharmonicity and electron-phonon
coupling.  The anharmonic linewidth can, thus, be determined from the
actual experimental linewidth and from the electron-phonon
linewidth obtained {\it e.g.} by calculations.  Finally, the
temperature dependence of the phonon energy and linewidth is usually
almost entirely due to the anharmonicity. 
The experimental determination of the temperature dependence of the
phonon energy and linewidth thus provide a further source of
information. 

Raman spectra of MgB$_2$
\cite{quilty1,quilty2,bohnen,hlinka,martinho,kunc,rafailov,cappelluti2} 
show a single very-broad peak corresponding to an excitation
with E$_{2g}$ symmetry at $75$ meV,
({\it i.e.} $15\%$ higher than the theoretical harmonic
frequency of $65$ meV). Furthermore, the linewidth of the peak has a
very strong temperature dependence.
These two facts have been interpreted by several authors 
\cite{liu,kortus,choi1,yildirim,boeri,bohnen} as 
a signature of strong anharmonicity.
The presence of strong anharmonic-effects was further supported by
ab-initio calculations based on the frozen-phonon model
\cite{liu,kortus,choi1,yildirim,boeri,bohnen,kunc}.
However, the results obtained with this model are questionable,
since the frozen-phonon approximation is an oversimplified model which does not
include all the terms of the  anharmonic perturbation theory
\cite{cardona,debernardi1,lang,lazzeri,Calandra_PhysicaC}.

Ab-initio calculations of anharmonic effects including
all the leading terms in anharmonic perturbation theory
( including scattering  between different wave-vectors and different 
phonon modes ) \cite{shukla,lazzeri} show that 
the phonon frequency shift of the E$_{2g}$ mode is only 5\% of its
harmonic phonon frequency at $\Gamma$. This is the result of a
partial cancellation between a large positive four-phonon 
scattering term ($\approx +10\%$) and a negative three-phonon scattering 
term ($\approx -5\%$). This is different from what is found
by the more-approximate frozen-phonon calculations, where the
four-phonon scattering term is dominant 
\cite{liu,kortus,choi1,yildirim,boeri,bohnen,kunc}. 
This difference should be reflected in the temperature dependence of
the Raman-peak energy, since the four-phonon scattering term generates
an increase in energy with temperature, while the sum of three- and
four-phonons term should be slightly
negative\cite{lazzeri,Calandra_PhysicaC}.  

Inelastic X-ray scattering (IXS) measurements  do not 
support the hypothesis of strong anharmonic effects.
In fact, the phonon dispersion measured by IXS\cite{shukla,baron}
is in good agreement with
harmonic-frequency calculations\cite{kong,shukla} and is consistent
with a small anharmonic shift along the $\Gamma$A direction.
In particular, the IXS E$_{2g}$ phonon frequency near $\Gamma$ is
65 meV, {\it i.e.} $\sim$10 meV smaller than the Raman peak.
Ab-initio calculations of the phonon linewidth
are in very good agreement with IXS measurements \cite{shukla}
and give an anharmonic contribution to the phonon linewidth which
is negligible with respect to the larger electron-phonon coupling
contribution. Moreover, calculations\cite{shukla} reveal a very weak
temperature-dependence of the phonon-linewidth between  50 K and 300 K.
Finally, the authors of Ref.\cite{baron} claim that the
phonon-linewidth of the E$_{2g}$ mode is independent from temperature
between 40 K and 300 K along the $\Gamma$A. If confirmed,
this last finding would establish that anharmonic effects are weak.
However, the moderate resolution and the fact that in Ref.\cite{baron} the
E$_{2g}$ mode is seen only as a shoulder of the nearby E$_{1u}$ mode
do not allow for a definitive conclusion.

In this work we present a high-resolution inelastic X-ray scattering 
and Raman study, in order to settle the debate on the presence 
of important anharmonic effects in MgB$_2$.
The two techniques are complementary
since inelastic X-ray scattering probes the phonon dispersion
throughout the entire Brillouin zone. 
However, at the $\Gamma$ point the elastic (Bragg) 
scattering usually dominates, therefore masking the phonon
contributions, as it is the case in MgB$_2$. 
On the contrary, Raman experiments probe excitations at
very small exchanged momentum, ${\bf q}\sim 0$. 
We show with high statistics and high q-resolution IXS measurements
that the E$_{2g}$ mode linewidth is independent of temperature along
$\Gamma$-A, and furthermore determine the dispersion along $\Gamma$-M with the
best resolution to date.
We investigate the behaviour of the Raman-peak energy and linewidth
as a function of temperature.
All our results are consistent with the presence of small
anharmonic effects in MgB$_2$. We, finally, show that, on the basis
of recent theoretical results\cite{calandra,cappelluti,Calandra_PhysicaC},
there is no contradiction between the
presence of a Raman-peak at 75 meV and the X-ray E$_{2g}$ phonon
frequency at 65 meV $near$ $\Gamma$.

The paper is structured as follows. In sec. \ref{sec:methods} 
we describe the methodological
details involved in inelastic X-ray scattering, Raman spectroscopy and
{\it ab-initio} simulations. 
In sec. \ref{sec:results} we present our main results and in sec.
\ref{sec:discussion} we comment on the results and discuss the main
implications for the interpretation of Raman data.

\section{\label{sec:methods} Methods}

\subsection{Experiments}

\subsubsection{Sample}
The crystal used in our experiment was grown at a pressure of
30-35 kbar.
A mixture of Mg and B was put into a BN container in a cubic anvil device.
The temperature was increased during one hour up to
$\rm 1700-1800^{\circ}$C, kept stable for 1-3 hours, and finally decreased
during 1-2 hours. As a result plate-like $\rm MgB_2$ crystals were
formed of which we used a sample of about 
$\rm 1.00 \times 0.45 \times 0.08~ mm^3$ 
with a measured in-plane mosaic of $\rm 0.011^{\circ}$.

\subsubsection{Inelastic X-ray Scattering}
The experiment was carried out at the Inelastic X-ray Scattering
Beamline II (ID28) at the European Synchrotron Radiation Facility in
Grenoble (France). The x-ray beam from the undulators was monochromatized
by a cryogenically cooled silicon (111) double
crystal monochromator, followed by a back-scattering monochromator,
operating at a Bragg angle of 89.98 degrees, and utilizing the Si (9 9 9) 
reflection order \cite{verbeni}. The back-scattered beam was then focused by a 
platinum-coated toroidal mirror,
which provided a focal spot at the sample position of 0.270
(horizontal) and 0.080 mm$^2$ FWHM at the sample position. 
The scattered photons were energy-analyzed by a 6.5m Rowland circle
five-crystal-spectrometer \cite {masciovecchio1,masciovecchio2}. 
The analysers operate at the same 
reflection order as the monochromator, corresponding to a 
wavelength $\lambda$ = 0.696782 \AA$^{-1}$ 
(photon energy: 17794 eV). 
The overall instrumental function has a pseudo-Voigt
line-shape with a width $\Delta$E of 3 meV (FWHM) as determined from
the elastic scattering  
of a Plexiglas sample (at a Q-transfer of 10 nm$^{-1}$, and T = 10 K).
The energy-analyzed photons are detected by a
Peltier-cooled silicon diode detector which has an intrinsic energy
resolution of 400 eV. The dark counts due to electronic and
environmental noise amounts to about 0.2 counts/minute. Further
components of the spectrometer are an entrance pinhole, slits in front
of the analysers in order to define the momentum transfer resolution
and a detector pinhole for aberrant ray suppression. 

The analyser crystals are temperature stabilized at 
295.65 K with a typical stability of 1 mK/24h. 
The momentum transfer $Q = 2~ k_i~ \sin(\frac{\theta}{2})$, where
$k_i$ is the incident photon wave vector and $\theta$, the scattering angle,
is selected by rotating the spectrometer around a vertical axis
passing through the scattering sample in the horizontal plane. 
Since there are five independent analyser systems, spectra at five different
momentum transfers can be recorded simultaneously. 
Their separation is energy dependent and for the Si(9 9 9)
reflection it amounts to 2.43 nm$^{-1}$. 
The energy scans are performed by varying the lattice spacing of the
monochromator temperature while the analyser temperature is kept fixed. 
Conversion from the temperature scale to the energy scale is accomplished by
using the known thermal expansion coefficient of silicon: 
$\alpha$(T) = $\alpha_0$ + $\beta$ (T-T$_0$)  with 
$\alpha_0$ = $2.581\times10^{-6}$ 1/K 
and $\beta$ = $0.016\times10^{-6}$ 1/K$^2$, T$_0$ = 295.65 K \cite{bergamin}. 

The validity of this conversion has been checked by comparing the
measured diamond dispersion curve for longitudinal acoustic and optic  
phonons with well established inelastic neutron scattering results. 

The energy scans were fitted using a sum of pseudo-Voigt:

\begin{equation}
I\left (
(1-\eta) \frac{\Gamma /2}{(\epsilon-\epsilon_0)^2+\Gamma^2 /4} + \eta  
\exp \left (-\frac{(\ln 2)(\epsilon-\epsilon_0)^2}{\Gamma^2 /4} \right )
\right ) 
\end{equation}

\noindent
and Lorentzian functions: 

\begin{equation}
I \frac{\Gamma /2}{(\epsilon-\epsilon_0)^2+\Gamma^2 /4} 
\end{equation}

\noindent
functions, where $\epsilon=\hbar\omega=h\nu$ is the
energy. Pseudo-Voigt functions were used to fit elastic and resolution-limited
inelastic contributions, with $\Gamma$ and $\eta$ parameters adapted to
match the instrumental function, while a Lorentzian line-shape was used to
fit the E$_{2g}$ mode, with a width $\Gamma$ fitted to the data.  
Alternatively, a convolution with the experimental resolution function has
been used. In this case, the response function was composed of a sum
of delta functions (for elastic and resolution-limited inelastic
contributions) plus a Lorentzian for the E$_{2g}$ mode. The two data
analysis procedures gave consistent results. 

The fitting algorithm used a $\chi^2$-minimisation routine \cite{james}, 
with the condition that the detailed balance between Stokes and anti-Stokes 
excitations is fulfilled. A constant background, coming essentially from 
electronic noise, was added.

The MgB$_2$ crystal was glued on a copper sample holder, which was mounted
inside a vacuum chamber on the cold finger of a closed-loop helium
cryostat (Cryomech ST15). The temperature was measured by a silicon
diode attached to the sample holder. 
An independent check of the sample temperature was provided by the
Stokes-Anti-stokes intensity ratio. 

\subsubsection{Raman measurements}

Raman spectra were excited with the 488\,nm line of an Ar laser
focused with an optical microscope (long distance 100x objective). The
scattered light was analyzed by a single-grating \textsc{Labram}
spectrometer equipped with a CCD. Temperature dependent measurements
were performed in an \textsc{Oxford} flow cryostat with a sapphire
window to preserve the polarization of the light. We used a 600\,1/nm
grating to record a wide spectral window (5000\,cm$^{-1}$ 
or 620 meV). 
The loss in resolution is irrelevant for MgB$_2$, because its Raman
features are at least several hundreds of wave numbers in width. 
Some experiments were also performed with 514\,nm and a triple-grating
spectrometer to compare with published studies. We report the
as-measured data normalized to the excitation energy and accumulation
time. Unless noted otherwise, the spectra were not scaled or shifted
along the $y$ axis.

To analyze the symmetry of the scattered light we recorded Raman
spectra in various backscattering geometries $k_i(e_i,e_s)k_s$, where
$k_i$ and $k_s$ indicate the direction of the incoming ($i$) and
scattered ($s$) light and $e_i$ and $e_s$ the polarization (Porto's
notation) \cite{cardona82}. For $\bm{k}_i\| \bm{k_s}\| c$ we used
backscattering normal to the surface; in the $\bm{k}_i\|\bm{k}_s\|a,b$
configurations the laser was focused onto the side of a MgB$_2$
crystal. The polarization of the incoming light was rotated by a
Fresnel rhomb. The polarization of the scattered light was chosen by a
$\lambda/2$ wave plate combined with an analyzer. We verified the
polarization-dependent measurements by recording spectra on the (111)
and (001) surface of silicon and on single crystals of graphite, which
has the same point group as MgB$_2$. We also ensured the internal
consistency of the Raman intensities; e.g., the spectra recorded in
$c(a,a)c$ and $b(a,a)b$ were identical in shape and intensity as
required by the Raman selection rules.

The frequency of the $\sim600\,$cm$^{-1}$ ($74$ meV) peak of
MgB$_2$ reported in the literature varies between $74$ and $77$ meV at
room temperature.\cite{quilty1,martinho,rafailov,chen,hlinka} Part of
the discrepancy might be due to a varying sample quality. However, a
major contribution comes from the uncertainty resulting from the
functional form used for the background and the peak. We tested several
fitting routines and found that the uncertainty for the position and
width of a Lorentzian is around 1.5\% or 1.2 meV. On the other hand,
when using a Fano line-shape the bare phonon frequency can be  
as low as 71 meV or as high as 79 meV (300\,K)
depending on the sign of the asymmetry parameter. The peak positions
and line widths reported in this paper were obtained by assuming the
600\,cm$^{-1}$ peak to be Lorentzian and the continuous background in
the $E_{2g}$ spectra to follow the same frequency dependence as the
$\alpha_{xx}(A_{1g})$ background signal (see Sect.~\ref{select}
for details). Other fit functions shift the position and width. We
also observed a change in the phonon frequency when reducing the
energy range in a fit. For our analysis we used the measured spectra
between 100\, and 2000\,cm$^{-1}$ ($\sim$ 10 and 250 meV). Smaller
windows yield a strong scattering in the peak positions. 
We stress, however, that varying fit routines resulted in a constant
offset of the temperature dependence or stronger scattering in the
data. 
The functional dependence of the Raman peak position and width on
temperature, however, remained unaffected. 

\subsection{Calculations}

Electronic structure calculations were performed using
density functional theory (DFT) and phonon frequencies were calculated 
in the framework of linear response theory \cite{DFPT} using the PWSCF/espresso
package\cite{PWSCF}. We used the generalized gradient approximation
\cite{PBE} and norm conserving pseudo-potentials \cite{Troullier}.
For $\rm Mg$ we used non-linear core corrections \cite{NLCC}
and we treated the $2s$ and $2p$ levels as core states.
The wave-functions were expanded in plane waves using a $35$ Ry cutoff.
The calculations were performed with the experimental
crystal structure, namely $a=3.083$~\AA~ and $c/a=1.142$.
For the phonon frequencies calculation,
we used a $16\times 16\times 16$ Monkhorst-Pack grid
for the electronic Brillouin Zone  integration and first order Hermite-Gaussian smearing 
\cite{Degironcsmear}
of $0.025$ Ry. 

The electron-phonon coupling contribution to the 
linewidth (full width half maximum ) at momentum ${\bf q}$
for the $\nu$ phonon mode, with frequency $\omega_{{\bf q}\nu}$
can be written as \cite{Allen}:

\begin{equation}
\gamma_{{\mathbf q}\nu} = \frac{4\pi \omega_{{\mathbf q}}}{N_{k}} \sum_{{\mathbf
k},n,m} |g_{{\mathbf k}n,{\mathbf k+q}m}^{\nu}|^2 \delta(\varepsilon_{{\mathbf k}n}) \delta(\varepsilon_{{\mathbf k+q}m})
\label{eq:linewidth}
\end{equation}

where the ${\bf k}$-sum is extended over the Brillouin Zone ,
$N_k$ is the number of $k$-points in the sum,
and $\varepsilon_{{\bf k}n}$ are the electronic energies, measured
with respect to the Fermi level, of the $n$ band at point ${\bf k}$.
The matrix element is
$g_{{\bf k}n,{\bf k+q}m}^{\nu}= \langle {\bf k}n|\delta V/\delta
e_{{\bf q}\nu} |{\bf k+q} m\rangle /\sqrt{2 \omega_{{\bf q}\nu}}$, 
where $\delta V/\delta e_{{\bf q}\nu}$ is
the derivative of the Kohn-Sham potential $V$ with respect to the phonon 
coordinate $e_{{\bf q}\nu}$ normalized over the unit cell.

The electron energies $\varepsilon_{{\mathbf k}n}$
and the $g_{{\bf k}n,{\bf k+q}m}^{\nu}$ matrix element
in eq. \ref{eq:linewidth} are obtained fully {\it ab initio}. The
actual sum over the Brillouin zone is calculated on a 
$N_k=27\times 27\times 27$ uniform 
k-point mesh shifted by a random vector with respect to the origin.

The x-ray structure factor for one-phonon scattering at temperature $T$,
frequency $\omega$ and reciprocal-space vector
${\bf q}+{\bf G}={\bf Q} $ (with
${\bf q}$ belonging to the first Brillouin Zone  and ${\bf G}$ a
reciprocal-lattice vector, and ${\bf Q}$ the total momentum transfer)
is obtained from 
\begin{equation}
S({\bf Q},\omega)=\sum_{\nu}G_{\nu}({\bf q},{\bf G}) F_{\nu}(\omega,T,{\bf G})
\end{equation}
with
\begin{eqnarray}
G_{\nu}({\bf q},{\bf G})&=&
\left|\sum_{\alpha} f_{\alpha}({\bf Q})
\frac{e^{-W_{\alpha}}}{\sqrt{M_{\alpha}}}
\left[({\bf Q})\cdot{\bf e_{{\bf q}\nu}^{\alpha}}\right]
e^{i{\bf Q}{\bf R_{\alpha}}}\right|^{2} \nonumber \\
F_{\nu}(\omega,T,{\bf G})&=&\sum_{\eta=-1}^{1}
\frac{\langle n_{{\bf q}\nu} \rangle
+\frac{1}{2}+\frac{\eta}{2}}{\omega_{{\bf q}\nu}}
\cdot\delta(\omega+\eta\omega_{{\bf q}\nu})
\end{eqnarray}
where $\alpha$ labels the atoms in the unit cell, $\nu$ the mode and  
$f_{\alpha}({\bf Q})$ is the atomic form factor 
calculated as in ~\cite{FormFactor}. 
Moreover, $M_{\alpha}$, ${\bf R}_{\alpha}$ and $W_\alpha$ 
are  the atomic mass, position in the unit cell and the Debye-Waller
factor, respectively.  
Finally,  $\langle n_{{\bf q}\nu} \rangle$ is
the phonon occupation number and ${\bf e_{{\bf
q}\nu}^{\alpha}}/\sqrt{M_{\alpha}}$ is the atomic displacement of the
atom $\alpha$ corresponding to the phonon eigenvector ${\bf e}_{{\bf q}\nu}$. 
The value $\eta=-1$ $(\eta=1)$ identifies Stokes (anti-Stokes) processes.

In what follows we neglect the Debye-Waller factor and we smear the
$\delta-$functions 
with the convolution between the phonon intrinsic lineshape calculated {\it ab
initio} and the instrumental function (3 meV FWHM).
For the calculation of the x-ray structure factor we used the atomic factors
for neutral Mg and B from \cite{FormFactor} and the phonon patterns calculated 
{\it ab initio}. 

\section{\label{sec:results} Results}

\subsection{\label{ixsres} Inelastic X-ray Scattering results}

The reciprocal points in the following section are expressed in terms
of hexagonal parameters $a=b=3.0828$, \AA~ $c=3.5186$ \AA, 
$\alpha=\beta=90^{\circ}$ and $\gamma=120^{\circ}$. In these
coordinates, all the measurements were recorded in the Brillouin Zone  centered
around (2,1,0). 

\begin{figure}
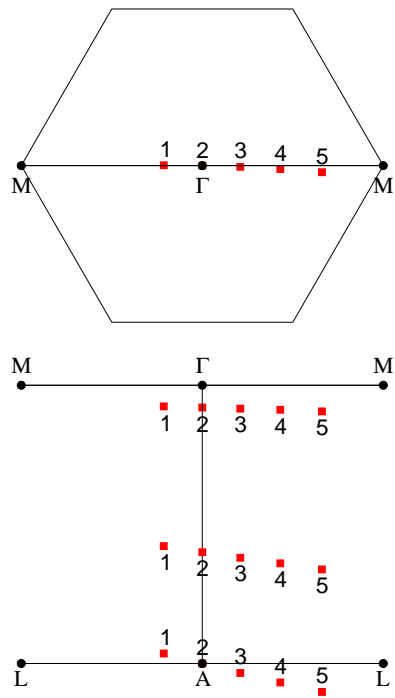

\includegraphics[scale=0.2,angle=0]{GApos-inplane.eps}\vspace{12pt}

\includegraphics[scale=0.2,angle=0]{GApos-ooplane.eps}
\caption{\label{GApos} (Color online) Positions of the analyzers (red
squares) for the measurements along 
$\Gamma$-A~(see Fig. \ref{temp}), and near $\Gamma$ (see Fig. \ref{Ghres}). 
$\Gamma$=G(HKL)=(2,1,0).\\}
\end{figure}

\begin{figure}
\includegraphics[scale=0.2,angle=0]{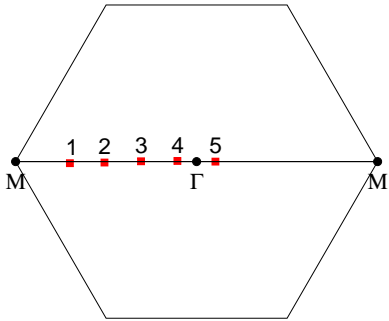}
\caption{\label{GMpos} (Color online) Positions of the analyzers (red
squares) for measurements along $\Gamma$-M~ (see 
Fig. \ref{dispersion}). $\Gamma$=G(HKL)=((2,1,0).\\} 
\end{figure}

In Fig. \ref{GApos} and \ref{GMpos} we show the relative position of
the analyzers for the measurements reported below. 

\subsubsection{Dispersion along $\Gamma$-M}

\begin{figure}
\includegraphics[scale=0.4,angle=0]{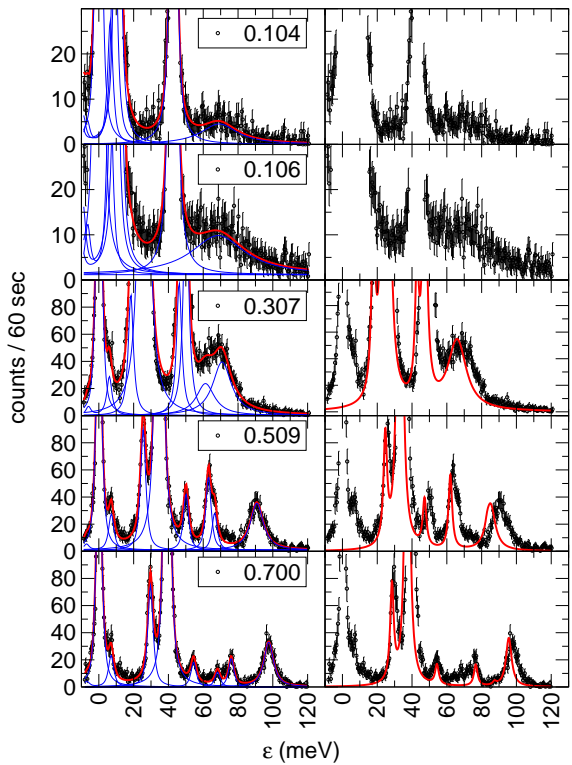}\hspace{48pt}
\caption{\label{scanGM} (Color online)
Left: IXS data (circle) and fit
  (lines) along $\Gamma$-M for T=46 K.
Right: IXS data (circle) and calculations
  (lines) along $\Gamma$-M for T=46 K.
 The legend box indicate the $\xi$ components of the $\mathbf{q}$
  reduced vector (see text).
\\}
\end{figure}

\begin{figure}
\includegraphics[scale=0.4,angle=0]{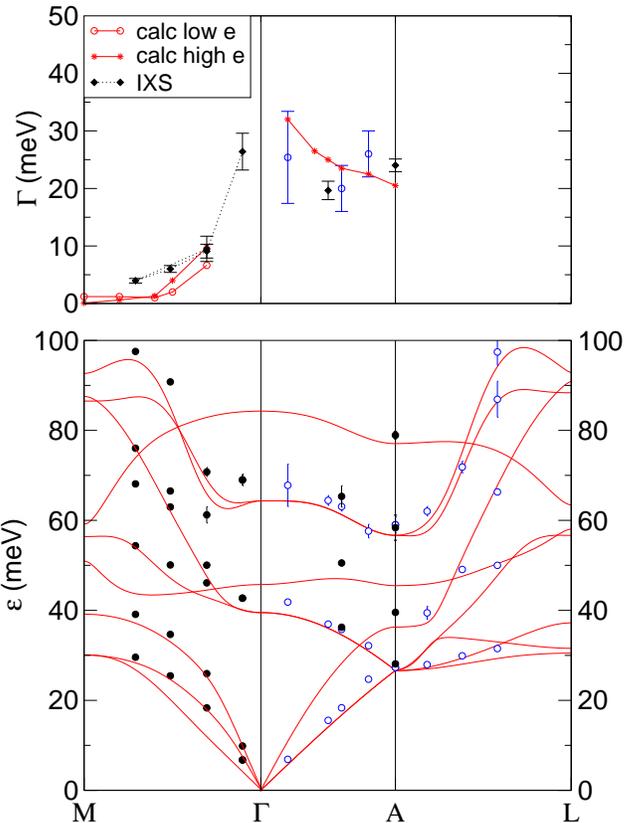}
\caption{\label{dispersion} (Color online)
Bottom panel: MgB$_2$ phonon dispersion. IXS data (circles) from
\cite{shukla} (blue open) and the present work (black full), compared
to calculations (red lines).
Top panel: E$_{2g}$ mode width from fit of the IXS data of the
present work (black diamonds) and Ref. \cite{shukla} (blue open with
errorbars) and calculations (red points with continuous line
connected). Stars corresponds to the high energy and open circles to
the low energy E$_{2g}$ modes when not degenerate. \\}
\end{figure}

The experimental results shown in figure \ref{scanGM} were obtained
exploiting the signal from all five analyzers. The relative Q
displacement between the analyzers is fixed by the spectrometer
geometry, and for longitudinal modes such a configuration can be close
to a high symmetry direction.   
This opportunity has been as well exploited in previous works on
MgB$_2$ \cite{shukla, baron}. Here the difference is that, using a
different incoming wavelength of 0.696782 \AA$^{-1}$, we have higher energy
resolution and larger steps in reciprocal space, corresponding to
the different fixed angular displacement of the analyzers.  
For a position close to the $\Gamma$-M line, 
this implies Q points G + q = $(2 , 1 , 0 ) + (\xi, \upsilon, \zeta)$ 
where $\upsilon$ and $\zeta$ are small compared to $\xi$. 
The configuration chosen for this experiment was as follows: 
\begin{itemize} 

\item[-]{analyzer 5: } Q = (2.052 1.000 0.0), $|\xi| \approx$ 0.052
  or 0.104 $\Gamma$-M;
\item[-]{analyzer 4: } Q = (1.948 1.002 0.0), $|\xi| \approx$ 0.053
  or 0.106 $\Gamma$-M ;
\item[-]{analyzer 3: } Q = (1.847 1.001 0.0), $|\xi| \approx$ 0.1535
  or 0.307 $\Gamma$-M;
\item[-]{analyzer 2: } Q = (1.744 0.997 0.0), $|\xi| \approx$ 0.2545
  or à.509 $\Gamma$-M;
\item[-]{analyzer 1: } Q = (1.645 0.990 0.0), $|\xi| \approx$ 0.350
  or 0.700 $\Gamma$-M;

\end{itemize}

 The positions in the hexagonal plane are shown in Fig. \ref{GMpos}. 
Note that analyzers 4 and 5 are placed symmetrically around the
chosen zone center. These positions are equivalent for phonon frequencies but
not for dynamical structure factors. Contribution along $<0, \upsilon,
0>$ are less than 0.005 r.l.u. for analyzer 5 to 2 and 0.01 r.l.u. for
analyzer 1. 

We measured the IXS spectra using a slit opening in front of the
analyzers of 60 mm in both horizontal and vertical direction, amounting to a Q
resolution of $\pm$0.416 nm$^{-1}$ for analyzers 1, 2 and 3, while
analyzers 4 and 5 had 60 $\times$ 20 mm$^2$ vertical $\times$
horizontal opening. 
This corresponds to a main contribution of $\pm$0.0022
r.l.u. in the  ( 0 , $\upsilon$ , 0 ) direction and to
$\pm$0.008 r.l.u. in the perpendicular ( 0 , 0 , $\zeta$ ) ( or
$\Gamma$-A) direction. 

In Fig. \ref{scanGM}, left panel, we show the IXS spectra
obtained at $\approx$  46 K for this configuration, together with the
fit used in order to extract mode frequencies as well as linewidth
values  for the E$_{2g}$ mode. 
A recent paper \cite{baronprb} suggests that
the MgB$_2$ IXS spectra show additional structures due to
two-phonon contributions. 
Although we do not have analyzed our data in
this way, such contributions may actually explain small deviations of
our fit results. In the right panel we show the same data
compared with \textit{ab-initio} calculated spectra, which correctly
reproduce the measurements. 

The same analysis is done for points along $\Gamma$-A (see
Sec. \ref{ixstemp}). These are grouped together with previous results
at lower resolution\cite{shukla} in order to experimentally determine the
phonon dispersion and the linewidths along all the high symmetry directions.  
We can thus compare the experimental phonon dispersion with 
the theoretical calculation along the high symmetry lines,
as shown in Fig. \ref{dispersion}, bottom. 
In the top panel, we show the experimental line-width 
$\Gamma$, for the E$_{2g}$ mode, de-convoluted by the instrumental
broadening, in comparison with the calculated linewidth.  
The agreement of the calculated frequencies with experiment is
remarkable. The same is observed for the agreement between the
experimental line-width,  
de-convoluted by the instrumental function, 
and the calculated electron-phonon-coupling contribution to the
intrinsic E$_{2g}$ linewidth.  
Note that data from Ref. \cite{shukla} are obtained at room
temperature and with 6.1 meV instrumental function FWHM. 

\subsubsection{\label{hrg} High resolution line-width  E$_{2g}$
measurements close to $\Gamma$} 

\begin{figure}
\includegraphics[scale=0.35,angle=0]{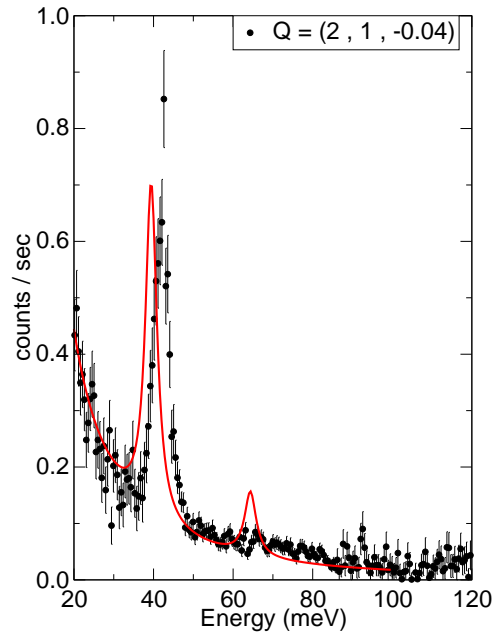}
\caption{\label{Ghres} (Color online) Room temperature IXS data at
 0.08$\times\Gamma$-A. 
 The line is an  ab-initio simulation convoluted with the energy resolution, 
 \textit{\textbf{without electron-phonon coupling}}
 broadening  effect on the E$_{2g}$ mode.}
\end{figure}

Fig. \ref{Ghres} shows the room temperature MgB$_2$ IXS spectra
for a Q-value corresponding to 0.08$\times\Gamma$-A, together with the
data, we show calculation convoluted with the instrumental resolution
for all modes, including the E$_{2g}$ one. The calculation was scaled as
in Fig. \ref{temp}.  

We measured the IXS spectra using a slit opening in front of the
analyzers of 20 mm in both horizontal and vertical direction,
equivalent to a Q resolution of $\pm$0.139 nm$^{-1}$. 
This corresponds to a main contribution of $\pm$0.0022 reduced lattice
units (r.l.u.) in the  ( 0 , $\upsilon$ , 0 ) direction and to
$\pm$0.0026 r.l.u. in the perpendicular ( 0 , 0 , $\zeta$ ) ( or $\Gamma$-M)
direction. 

\subsubsection{\label{ixstemp} Temperature-dependence of the IXS spectra}

\begin{figure}
\includegraphics[scale=0.4,angle=0]{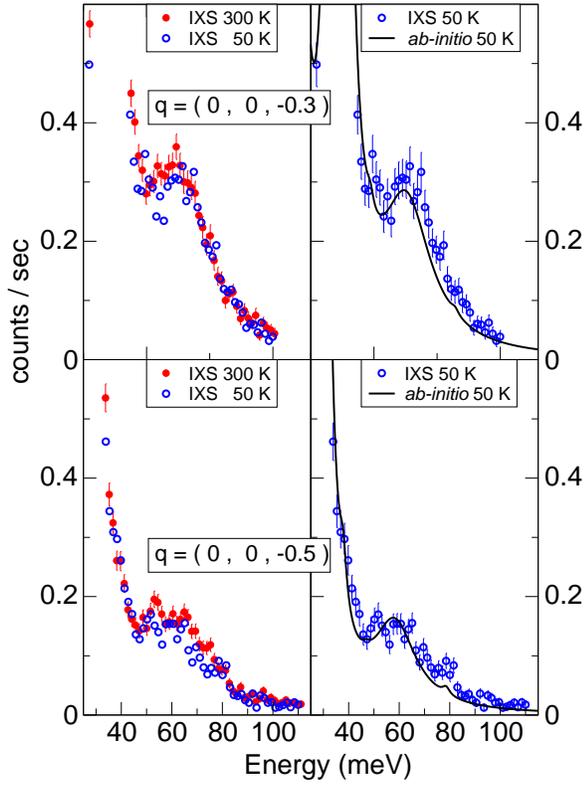}
\caption{\label{temp} (Color online) Left: IXS data at T=300 K (red
full points, with errorbars) and T=46 K (blue empty points). 
Right: IXS data at T=46 K (blue empty points) and ab-initio simulation (lines)
 Measured spectra and calculation spectra at 0.6$\times\Gamma$-A
 (top) and at the A point (bottom).\\}
\end{figure}

In figure \ref{temp} we compare IXS E$_{2g}$  spectra measurements and
calculations performed at two different sample temperatures ($\sim$ 50
K and 300 K) for Q=(HKL) corresponding to a phonon propagation vector of 
(0, 0, 0.3) (top panel) (\textit{i.e.} at 60\% of the
$\Gamma$-A line) and the zone boundary A
point ( 0 , 0 , 0.5) (bottom panel).

To induce a maximum effect from phonon-phonon
scattering (anharmonicity) \cite{shukla, lazzeri}, we choose the
largest possible temperature difference in the normal-state region,
\textit{i.e.~} from room temperature to T above
$\sim$ 40 K. This was done to avoid effects due to the superconducting gap.   
More precisely, the measured temperature by the probe on the sample
holder was stable at 46 K and 298 K respectively.  

We measured the IXS spectra using a slit opening in front of the
analyzers of 60 mm in both horizontal and vertical direction,
equivalent to a Q resolution of $\pm$0.416 nm$^{-1}$. 
This corresponds to a main contribution of $\pm$0.0065 r.l.u. along
the ( 0 , $\upsilon$ , 0 ) direction and to 
$\pm$0.008 r.l.u. along the perpendicular ( 0 , 0 , $\zeta$ )
direction (\textit{i.e.~} along $\Gamma$-A).
The theoretical spectra are arbitrarily normalized to reproduce the
height of the highest vibrational peak (E$_{1u}$, not shown in
Fig.~\ref{temp}.)  

\subsection{Raman results}

\subsubsection{\label{select} Raman selection rules}

\begin{table}
\caption[]{Raman selection rules for the $D_{6h}$ point group of
MgB$_2$. $a$ and $b$ are vectors within the hexagonal plane forming an
angle of $90^\circ$. Since the hexagonal lattice is isotropic in the
plane $a$ and $b$ can be chosen arbitrarily (with respect to the
lattice vectors). $c$ is normal to the plane. The $\alpha_{ii}$ denote
the element $ii$ of the corresponding Raman
tensor.\label{tab_selection}} 
\begin{ruledtabular}
\begin{tabular}{ccc}
configuration&Raman intensity&symmetry\\\hline
$c(a,a)c,\,b(a,a)b$&$\alpha_{xx}^2+\alpha_{yy}^2$&$A_{1g}, E_{2g}$\\
$c(a,b)c,\,c(b,a)c$&$\alpha_{xy}^2$&$E_{2g}$\\
$b(c,c)b$&$\alpha_{zz}^2$&$A_{1g}$\\
$b(a,c)b,\,b(c,a)b$&$\alpha_{xz}^2$&$E_{1g}$
\end{tabular}
\end{ruledtabular}
\end{table}

\begin{figure}
\centerline{\includegraphics[width=70mm]{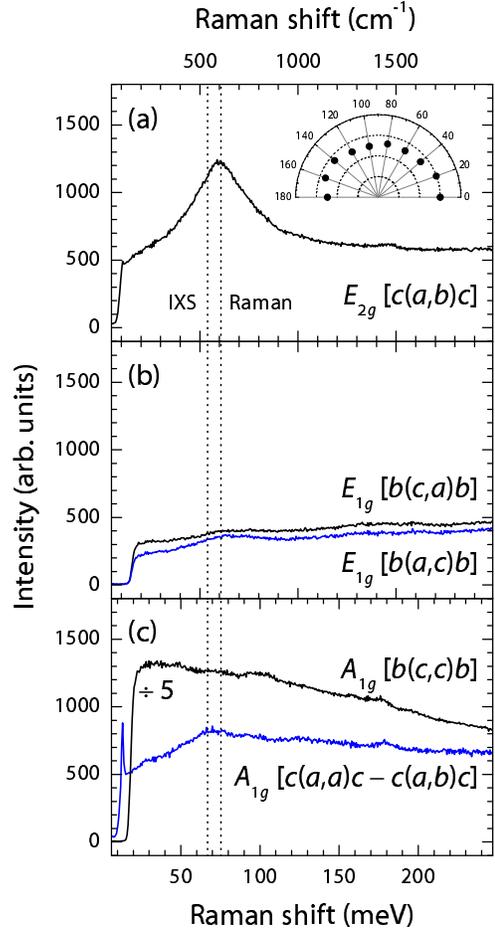}}
\caption[]{(Color online) Raman selection rules and symmetry of the
scattered light in MgB$_2$ (room temperature). (a) $E_{2g}$ component
with the peak at 75 meV (608\,cm$^{-1}$). 
The background measured under crossed
polarization is also of $E_{2g}$ symmetry in contrast to parallel
polarization, where it has $E_{2g}$ and $A_{1g}$ [see (c)]
contribution. 
The inset shows the constant integrated intensity of the 75 meV peak
when rotating the in-plane polarization. 
(b) $E_{1g}$ symmetry obtained in two scattering configurations. 
(c) $\alpha_{zz}$ and $\alpha_{xx}=\alpha_{yy}$ $A_{1g}$ 
component of the scattering intensity. 
The latter was obtained by subtracting the $c(a,a)c$ and
$c(a,b)c$ Raman spectra. The two dashed lines mark the phonon
frequency of the $E_{2g}$ peak measured with Raman and inelastic X-ray
scattering.}  
\label{selection}
\end{figure} 

Table~\ref{tab_selection} gives the selection rules for the Raman
active representations of MgB$_2$ ($D_{6h}$ point group). Note that
the intensity of $E_{2g}$ is independent of polarization as long as
the light incidents and scatters normal to the $ab$ plane. In
Fig.~\ref{selection} we present the four independent components of the
Raman scattered light from the MgB$_2$ crystal. The
$\sim75$ meV peak is indeed only found in the $E_{2g}$
scattering configuration,\cite{quilty1,hlinka}
Fig.~\ref{selection}(a), and its intensity is independent of the
in-plane orientation of the crystal (see inset). The weak feature at
75 meV in the $E_{1g}$ $b(a,c)b$ configuration is within
experimental error. The two dotted lines mark the $E_{2g}$ peak
frequency as obtained by Raman (75 meV) and inelastic X-ray
scattering ($\approx69$ meV) at 0.1 $\Gamma$-M~ and \emph{ab-initio}
calculations at $\Gamma$ (70 meV) \cite{lazzeri}. As noted earlier, the two
measurements give phonon frequencies differing by at least $\sim
6\,$meV. 

Our Raman spectra, consistently with the literature
\cite{quilty1,quilty2,bohnen,hlinka,martinho,kunc,rafailov,cappelluti2},
are smooth. They do not show any structure that can be interpreted as a
multiple-phonon peak, as proposed in Ref.~\onlinecite{baronprb}.
In general, two-phonon Raman spectra have more than 
one symmetry component \cite{cardona82}. However the $A_{1g}$ and
$E_{1g}$ spectral contributions in Fig.~\ref{selection} do not show
any sharp feature. 

The polarization dependence of the continuous background for
$\alpha_{xz}(E_{1g})$, Fig.~\ref{selection}(b), and
$\alpha_{zz}(A_{1g})$, Fig.~\ref{selection}(c), agree with
Quilty~\emph{et al.}\cite{quilty1}. The scattering intensity increases
linearly with the frequency of the excitation. We extracted the
$\alpha_{xx}(A_{1g})$ background component from the measured spectra as
shown in Fig.~\ref{selection}(c). In contrast to the frequency
dependence of the $zz$ and $xz$ component of the Raman tensor,
$\alpha_{xx}$ has a maximum at $\sim60\,$meV and decreases in
intensity towards smaller frequencies. The $\alpha_{xy}(E_{2g})$
background component appears to be similar to $\alpha_{xx}(A_{1g})$;
the overall intensity is almost identical and the decrease in
background intensity towards small frequencies is present as well as
can be seen in Fig.~\ref{selection}(a). This points towards a common
origin of the electronic $E_{2g}$ and $A_{1g}$ scattering, e.g., such
a behavior might arise from a coupling of electrons that are
correlated with the $E_{1g}$ representation of the MgB$_2$ point
group. 

The precise origin of the electronic background in the MgB$_2$ Raman
spectra is not known. It is remarkable that the signal is strong and
approximately constant over a wide energy range (at least up to
600 meV the highest energy in our measurements).

\subsubsection{Temperature-dependence of the Raman spectra}

\begin{figure}
\centerline{\includegraphics[width=80mm]{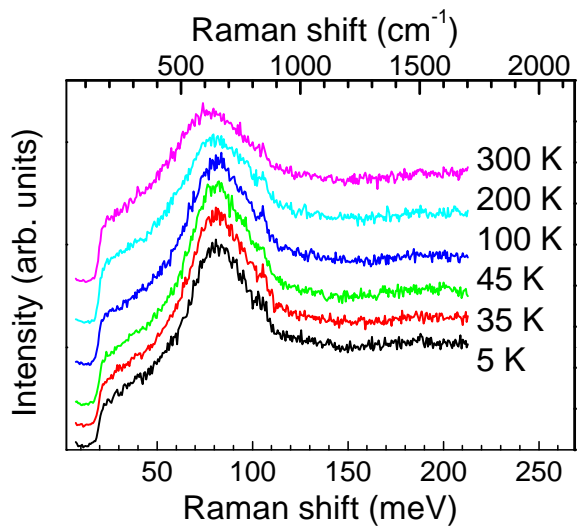}}
\caption[]{(Color online) E$_{2g}$ Raman spectra (raw data) of MgB$_2$
between 5\,K and room temperature in $c(a,b)c$ scattering configuration. 
An offset has been added  for clarity (see the zero line
at the cut-off of the notch filter).  } 
\label{temperature}
\end{figure}

\begin{figure}
\centerline{\includegraphics[width=80mm]{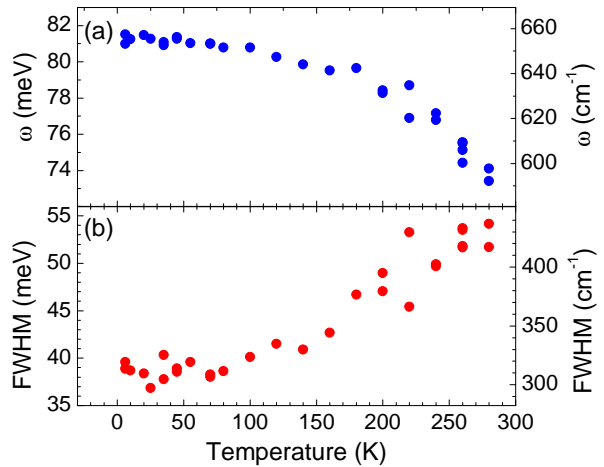}}
\caption[]{(Color online) Position (a) and full width at half maximum
(b) for the 75 meV (600\,cm$^{-1}$) Raman peak. 
The spectra were fitted by a Lorentzian
plus a background. The frequency dependence of the background is assumed
to be the same as the
$\alpha_{xx}(A_{1g})$ component, see Fig.~\ref{selection}(c).} 
\label{pos}
\end{figure}

Figure~\ref{temperature} shows Raman spectra measured between 5\,K
(bottom) and room temperature (top). The 75 meV (600\,cm$^{-1}$) peak decreases
in frequency and broadens at higher temperatures. The frequency shift
is more pronounced than reported by Quilty~\emph{et al.}\cite{quilty1}
and disagrees with the finding by Martinho~\emph{et
al.}\cite{martinho}, who reported a constant peak frequency between 10
and 300\,K. The shift in the peak maximum agrees with
Rafailov~\emph{et al.}\cite{rafailov}, although this is somewhat
difficult to judge given the narrow Raman frequency window in the
reference. Surprisingly, we find the electronic background to be
independent of temperature. The intensity is constant (the spectra in
Fig.~\ref{temperature} are raw data) and there is no increase in the
scattering intensity at low frequency for low $T$ \cite{quilty1,chen}. 

The frequency of the $\sim75\,$meV (600\,cm$ ^{-1}$) Raman peak
decreases from 81 meV to 75 meV between 5\,K and room temperature as
we show in Fig.~\ref{pos}(a). This is accompanied by an increase in the line
width by 45\% or from 37 to 54 meV (Fig.~\ref{pos}(b)).

\section{\label{sec:discussion} Discussion}

\begin{table*}
\begin{ruledtabular}
\begin{tabular}{c|cc|cc|ccc}
&\multicolumn{2}{c|}{$\xi=0.0$}&\multicolumn{2}{c|}{$\xi=0.3$}&\multicolumn{3}{c}{$\xi=0.5$}\\
\hline
& Raman & Theory & IXS &Theory & IXS & Theory & Theory\\
&       & (Anharm.) &    & (EPC) &     &   (EPC)&  (Anharm.) \\
\hline
T=300K & 53 $\pm$ 3 & 1.21 & 23 $\pm$ 2 & 23.0 & 26 $\pm$ 2 & 20.35 & 2.13 \\
T=50K & 39 $\pm$ 2 & 0.16 & 20 $\pm$ 2 & 23.0 & 24 $\pm$ 2 & 20.35 & 0.10 \\
\end{tabular}
\end{ruledtabular}
\caption{\label{expvsthwidth} 
Measured (Raman and IXS) and calculated phonon-linewidth (FWHM) of the E$_{2g}$
phonon mode at the $(0,0,\xi)$ wave-vector in meV. 
Both the electron-phonon coupling (EPC) and the anharmonic contributions
(Anharm.) to the phonon linewidth are reported. 
Computational details are explained in
Refs. \cite{shukla,lazzeri}. Note that the anharmonic 
contribution has been calculated only at selected high symmetry point:
$\Gamma$ ($\xi=0.0$) and A ($\xi=0.5$). There is no EPC contribution
at $\Gamma$\cite{calandra}.} 
\label{tabella}
\end{table*}

The difference between the 75 meV Raman peak and the calculated
harmonic E$_{2g}$ phonon frequency (65 meV) has been attributed to the
presence of anharmonic effects which are supposed to be large
\cite{liu} along the entire $\Gamma$A direction. Here, we 
have determined using high resolution IXS the phonon dispersion and the
phonon linewidth along three high symmetry directions in the Brillouin
zone to verify whether anharmonic effects are relevant or not.  As
already shown in Refs.~\cite{shukla,baron}, the IXS phonon dispersion
is in good agreement with the calculated harmonic phonon dispersion.
To further determine the magnitude of anharmonic effects, in the
present work, we have also measured the phonon frequency shift and
linewidth as a function of temperature using Raman spectroscopy and
inelastic X-ray scattering.

Our Raman spectra show a well defined peak at $\approx 75$ meV having
E$_{2g}$ symmetry.  The energy of the 75 meV peak shifts downward and
broadens by increasing the temperature.  The frequency shift is more
pronounced than reported by Quilty~\emph{et al.}\cite{quilty1} and
disagrees with the finding by Martinho~\emph{et al.}\cite{martinho},
who reported a constant peak frequency between 10 and 300\,K.  The
decrease in energy of the Raman-peak on temperature increase is
not consistent with the claim that anharmonic four-phonon scattering
is the dominant anharmonic contribution
\cite{liu,kortus,choi1,yildirim,boeri,bohnen}.
Indeed, since the four-phonon scattering term is positive at $\Gamma$
\cite{liu,kortus,choi1,yildirim,boeri,bohnen,lazzeri},
the corresponding anharmonic shift should increase with
temperature (see {\it e.g} Eq.~4 in Ref.~\cite{lazzeri} ).  However,
even including a negative third order term, state-of-the-art
calculations\cite{lazzeri} report a substantially smaller decrease of
the E$_{2g}$ phonon frequency than what is presently measured.  
{\it This
suggests that anharmonicity is not the main mechanism determining the
temperature behavior of the Raman-peak energy and that the Raman-peak
is not due to a bare phonon-vibration}.
We also remark that the variation of the Raman shift with the temperature
(Fig.~\ref{temperature}) cannot be explained by a two-phonon 
contribution as proposed by Ref.\onlinecite{baronprb}
(see \textit{e.g.} Ref. \onlinecite{cardona82}).

The large increase of the Raman linewidth between 40 K and 300 K has
no equivalent in the IXS spectra.  Using high-resolution IXS we have shown
that the phonon linewidth along the $\Gamma$A direction is essentially
temperature-independent in the 46-300 K range. This was already
claimed in a previous paper \cite{baron}.  However in Ref.\cite{baron}
the E$_{2g}$ peak was only detected as a shoulder of the nearby
E$_{1u}$ mode, while in the present work we have a very 
well resolved peak for the E$_{2g}$ mode in the region from 0.5$\Gamma$A to
$A$.  As shown in Table \ref{expvsthwidth} and Fig. \ref{dispersion} (top
panel), {\it ab-initio} simulations are successful in predicting the phonon
linewidth (including electron-phonon and anharmonic damping) of
MgB$_2$ along all the high-symmetry directions considered.  This very
good agreement confirms that, in the region from 0.5$\Gamma$A to $A$,
phonon-damping is almost exclusively due to electron-phonon coupling
and not to anharmonic effects\cite{shukla,lazzeri}.

We finally want to stress that there is no contradiction between the
Raman-peak at 75 meV \textit{at} $\Gamma$ and the X-ray E$_{2g}$ phonon
frequency at 65 meV $near$ $\Gamma$ and {\it ab-initio} harmonic 
calculations (giving 65 meV for the E$_{2g}$, $\Gamma$ phonon).
The key to understand these differences is the presence of relevant 
dynamical and electronic effects in the phonon self-energy, as shown
in Ref. \cite{calandra,Calandra_PhysicaC,cappelluti,lazzeri_dynamic}.
So far, ab-initio phonon calculations in MgB$_2$ have been performed
within the adiabatic  approximation.
The adiabatic approximation assumes that the phonon is a {\it static}
perturbation, but, actually, a phonon is {\it dynamic} perturbation,
oscillating at the frequency $\omega$.
Within the adiabatic approximation, the phonon self-energy does not depend
on $\omega$, while, in general, the phonon self-energy {\it depends}
on $\omega$.
The adiabatic approximation is valid if the interatomic force constants
are istantaneous, that is the force on a given atom at a given time
depends on the position of the other atoms at the same time~\cite{allen80}.
In reality, however, the force on a given atom at a give time $t$ depends
on the position of the other atoms at a previous time $t'$~\cite{allen80}.
This dynamical effect can be treated within time-dependent perturbation
theory and can be relevant for zone-center optical phonons.

As shown in Refs.\cite{calandra,Calandra_PhysicaC}, the
E$_{2g}$ phonon frequency varies strongly for ${\bf q}$ near $\Gamma$
due to dynamical effects in the real-part of the phonon self-energy.
As an example, along the $\Gamma$M direction, the E$_{2g}$ phonon
frequency has a smooth behavior for ${\bf q}>0.1\Gamma$M and
increases abruptly for ${\bf q}<0.05\Gamma$M
(see fig. 3 in Refs.
\cite{Calandra_PhysicaC} or fig. 3(a) of Ref. \cite{calandra}).
The E$_{2g}$ X-ray dynamical structure factor is strong enough for accurate
statistics only for $q > 0.1 \Gamma$M, that is IXS are probing
the smooth region.
On the contrary, Raman spectroscopy probes excitations at much smaller
${\bf q}$ (${\bf q}<0.05\Gamma$M), i.e. in the region where the phonon
frequency is higher.
Concluding, the $\approx 10$ meV difference between the Raman and X-ray
peaks is expected from the theoretical findings of Refs.
\cite{calandra,Calandra_PhysicaC} and is 
due to the pathological behavior of the phonon self-energy at
$\Gamma$.

The disagreement between Raman data and the harmonic {\it ab-initio}
calculations, published so far, is due to the fact that these calculations
have been performed within the adiabatic approximation. 
When dynamical effects are neglected, the mentioned pathological
behavior is absent \cite{Calandra_PhysicaC,lazzeri_dynamic} and the
phonon dispersion is smooth also in the region near $\Gamma$. Therefore,
the calculated E$_{2g}$ frequency at $\Gamma$ 
is smaller than the actual frequency, which should be computed including
dynamical effects. These dynamical effects have no 
influence outside a small region around $\Gamma$ and, for this reason,
the {\it ab-initio} harmonic calculations done outside this region 
are correct and reproduce the  phonon dispersion. 

Finally, in Ref.\cite{calandra} it was shown that the Allen
formula\cite{Allen} gives a zero linewidth for the E$_{2g}$ mode at
$\Gamma$. As shown in fig.\ref{Ghres} this is not the case.  To
explain the presence of a non-zero linewidth at $\Gamma$, it is
necessary to include electron self-energy effects in the phonon
self-energy as pointed out by Cappelluti in Ref. \cite{cappelluti}.
According to Ref. \cite{cappelluti} these effects are responsible for
the unexplained softening of the Raman-peak energy and increase of the
Raman-peak linewidth by increasing the temperature.  However, the
model of Ref. \cite{cappelluti} is simplified, therefore giving only a
qualitative trend. More theoretical work is required to achieve
quantitative agreement with experiments.

\section{Conclusions}

We have presented a high-resolution inelastic X-ray scattering and
Raman study, in order to settle the debate on the presence or not of
important anharmonic effects in MgB$_2$.  First, we have shown with
high statistics and high q resolution measurements that the E$_{2g}$
mode linewidth is independent of temperature along $\Gamma$A, ruling
out a major contribution of anharmonicity in the lattice dynamics of
MgB$_2$. Moreover we measured the dispersion along $\Gamma$-M with the
best resolution to date. The E$_{2g}$ mode shows, along this line,
large variation of both phonon frequencies and line-width, as
previously observed along A-L \cite{shukla} and $\Gamma$-M\cite{baron}
directions.  Second, we have investigated the behavior of the
Raman-peak energy and linewidth as a function of temperature.  The
Raman-peak energy decreases as a function of increasing temperature.
This behavior is not reproduced by anharmonic ab-initio calculations
of the E$_{2g}$ $\Gamma$ mode (calculations done both with the
oversimplified frozen-phonon approach
\cite{liu,kortus,choi1,yildirim,boeri,bohnen,lazzeri}
or with state-of-the-art perturbation theory \cite{lazzeri}).  This
finding suggests that anharmonicity is not the main mechanism
determining the temperature behavior of the Raman-peak energy and
that the Raman-peak is not due to a bare phonon-vibration.  Finally,
we have shown that, on the basis of recent theoretical
results\cite{calandra,cappelluti,Calandra_PhysicaC}, there is no
contradiction between the presence of a 75 meV Raman-peak at $\Gamma$ 
and the X-ray E$_{2g}$ phonon frequency at 65 meV \textit{near} $\Gamma$.

In conclusion, the present results indicate that anharmonicity 
plays a marginal role in MgB$_2$. As a consequence, the explanation of
the reduced isotope effect\cite{choi1,choi2} - one of the most 
important unresolved issues in the physics of MgB$_2$ -  needs
to be reconsidered. 
At present, the explanation of the reduced isotope effect is one of the most
important unresolved issues in the physics of MgB$_2$.

\begin{acknowledgments}
We wish to acknowledge the ESRF for the support of experiment
HS-2598.
The Raman measurements were performed at the Technische Universit\"at 
Berlin. We thank C. Thomsen for kind hospitality in his lab, the use
of the Raman equipment, and helpful discussions. We are grateful to S.
Bahrs and D. Heinrich for help with the temperature-dependent Raman
measurements. We thank A. Ferrari for discussion on the
Raman measurements.  
Calculations were performed at the IDRIS supercomputing center 
(project 071202). 
This work was supported by the Swiss National Science Foundation 
through NCCR MaNEP. 
\end{acknowledgments}

\end{document}